\documentclass[aps,prb,twocolumn,showpacs]{revtex4-1}
\usepackage{graphicx}
\usepackage{amsmath}
\usepackage{amssymb}
\usepackage{color}

\begin{document}
	
	\title{How does the Phase Slip in a Current-Biased Narrow Superconducting Strip?}
	\author{Yu.N.Ovchinnikov}
	\affiliation{Max-Plank Institute for Physics of Complex Systems, Dresden, D-01187,
		Germany } 
	\affiliation{Landau Institute for Theoretical Physics, RAS, Chernogolovka, Moscow
		District, 142432 Russia}
	
	\author{A.A.Varlamov}
	\affiliation{CNR-SPIN, Viale del Politecnico 1, I-00133, Rome, Italy} 
	\affiliation{Materials Science Division, Argonne National Laboratory,
		9700 S. Cass Avenue, Argonne, Illinois 60639, USA}
	
	\date{\today}
	
	\begin{abstract}
		The theory of current transport in a narrow superconducting strip is revisited
		taking the effect of thermal fluctuations into account. The value of voltage drop
		across the sample is found as a function of temperature (close to the
		transition temperature, $T-T_{\mathrm{c}}$ $\ll T_{\mathrm{c}}$) and bias
		current $J<J_{\mathrm{c}}$ ( $J_{\mathrm{c}}$ is the critical current
		calculated in the framework of the BCS approximation, neglecting thermal
		fluctuations). It is shown that careful analysis of vortices crossing
		the strip results in considerable increase of the activation energy.
	\end{abstract}
	
	\maketitle

\section{Introduction}

The fundamental property of currents flowing dissipationless through
superconducting components is the underlying principle for the operation of numerous nano-electronic
devices. One component of particular interest is a narrow superconducting strip (NSS), in which thermal and quantum fluctuations can result in a resistive
state of the system. Understanding the role of such fluctuations is a problem of great importance. 
Various models have been proposed to explain the appearance of non-zero resistances in NSSs and its
temperature dependence in the region of low temperatures (for a review see
Refs.~\cite{S09,B13}).

The role of thermal fluctuations responsible for energy dissipation, when currents flow through a one-dimensional superconductor 
was considered for the first time in the paper by Langer and Ambegaokar \cite{LA67} almost
fifty years ago. The publication of this paper has strongly influenced all
subsequent studies in this field, becoming part of multiple monographs and handbooks on
superconductivity \cite{T75,A88,LV04}.

It is necessary to mention that a ``one-dimensional superconductor'' is
\textit{de facto} often  a narrow strip with finite width $L$, much less
than the Ginzburg-Landau coherence length $\xi_{\mathrm{GL}}\left( T\right)
= \sqrt{\pi\mathcal{D}/16T\tau}$ ($\tau=1-T/T_{\mathrm{c}}$ is the reduced
temperature and $\mathcal{D}$ is the electron diffusion coefficient \cite{G59}
). The energy dissipation in this system is related to phase-slip processes, i.e.,
the process of vortices/flux quanta crossing the strip. 
It is clear, that such events cannot be realized in the framework of a purely
one-dimensional model. Indeed, the solution found in Ref.~\cite{LA67} shows
that even when the current density flowing through the one-dimensional
superconductor reaches its critical value $J_{\mathrm{c}}$, the
minimal value of the order parameter is $\left( 2/3\right) ^{1/2}\Delta_{ 
\mathrm{BCS}},$ while in order to perform a phase slip event it should
become zero at least in one point. 

In this work we will resolve the mentioned paradox, describing the true
mechanism of phase-slip events in NSS and determining the corresponding
value of the activation energy. We will demonstrate that the saddle point
solution of the Ginzburg-Landau (GL) equation for the order parameter $%
\widetilde{\Delta}$ in presence of a fixed current $J,$ possessing at least
one vortex, exists only for weak enough currents $J<J_{\mathrm{c1}
}=\eta\left( L/\xi_{\mathrm{GL}}\right) J_{\mathrm{c}},$ ($J_{\mathrm{c}}$
is the critical current of the strip, and $\eta=0.0312$ is a small number
which will be found below). Under the expression ``saddle point solution'' 
$\widetilde{ \Delta}\left( x,y,J,r_{1},..r_{i},.\right) $ we understand the
solution of the GL equation, which depends not only on the coordinates $x,y,$
and current $J$, but also on some set of parameters $\{\mathbf{r}_{i}\}$
satisfying the extremal conditions for the GL functional $\mathcal{F}_{s}$: 
\begin{equation}
\frac{\partial}{\partial \mathbf{r}_{i}}\mathcal{F}_{s}\left[ \widetilde{%
\Delta } \left( x,y,J, \mathbf{r}_{1},.. \mathbf{r}_{i}\right) ,J\right] =0.
\label{GLSS}
\end{equation}
In the case under consideration, when one or several vortices penetrate the system
through its edge, those parameters can be chosen as the vortex center
coordinates (zeros of the order parameter function).

When the current $J$ exceeds the value $J_{\mathrm{c1}}$ the saddle point
solutions (\ref{GLSS}) leading to phase-slip events cease to exist and
the scenario described above does not hold anymore. In that case another mechanism comes into
play. In order to explain this, let us recall that the minimum of the GL free
energy is reached for the ground state, corresponding to a solution with
spatially independent modulus $|\Delta _{\mathrm{gs}}\left( x,y\right)
|=\Delta _{0}.$ When $J<J_{\mathrm{c1}}$ the saddle point solutions of the GL
equations, including vortices, exist with energies higher than the one of
the ground state. The transition from the ground-state to the saddle point
solution can be imagined as the motion of the order parameter
\textquotedblleft vector\textquotedblright\ in Hilbert space, accompanied by
the motion of the \textquotedblleft points\textquotedblright $\{\mathbf{r}%
_{i}\}$ in the finite-dimensional space of those parameters.

As we already said, in the case of \textquotedblleft strong \
currents\textquotedblright\ $J_{\mathrm{c1}}<J<J_{\mathrm{c}}$\ the saddle
point solutions of the GL equations, possessing vortices, do not exist anymore. In
this interval the minimal activation energy is reached at some function $%
\Delta _{\mathrm{v}}\left( x,y,J,\mathbf{r}_{1}\right) $\ corresponding to
the state with a single vortex. We choose such a gauge (i.e. the form of
vector potential $A$) where the phase of the order parameter is determined
 by the vortex position only and the boundary conditions at the strip edges.
The modulus of $\Delta _{\mathrm{v}}\left( x,y,J,\mathbf{r}_{1}\right) $\ is an even function of the longitudinal coordinate $y$ in
that case.

\begin{figure}[th]
\includegraphics[width=0.7\columnwidth]{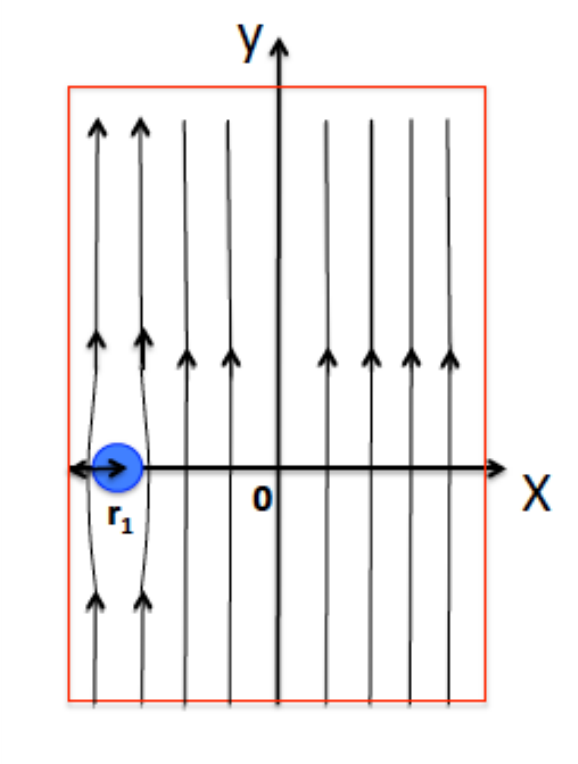}
\caption{Distribution of the current flowing in the strip in presence of
a vortex located in distance $r_{1}$ from the edge of the strip.}
\label{current}
\end{figure}

In order to determine the order parameter in the state with vortices and subsequently to calculated the corresponding value of free energy, we will use the variational principle with respect to several free parameters in the following. One of them is
the distance $r_{1}$\ from the edge of the strip to the center of a vortex
(i.e. the coordinates of the vortex center are $x_{1}=-L/2+r_{1}$\ and $%
y_{1}=0,$\ see Fig. \ref{current}). We will look for the maximum value $%
r_{1}^{\mathrm{ext}}$\ for which the conditional extremum of the free energy
functional (i.e. the extremum at given value of the parameter\ $r_{1}$)
still exists. If the vortex penetrates further into the system, i.e., for $r_{1}>r_{1}^{%
\mathrm{ext}}$, such extremum ceases to exist. 
Let us recall that the order parameter of the current-biased one-dimensional
superconducting channel, which corresponds to the saddle point solution of
the GL equation does not have zeros at all \cite{LA67}. 
When we consider a strip with finite width instead of such a channel a lateral penetration of a
vortex is possible. This allows to suppress the modulus of the
order parameter to zero at some point allowing a phase-slip event at this location. 
It is clear that such a deformation of the order parameter on a small scale requires some
excess energy. The system partially compensates this energy loss by means of a deformation of the order parameter distribution in relatively large
distances from the vortex along the strip with respect to corresponding
one-dimensional solution \cite{LA67}. This deformation is accounted for by means
of the variational parameter $y_{0}$. We derive the equations which
allow to determine the value of the parameter $y_{0}$\ maximizing $r_{1}$
(i.e. $r_{1}^{\mathrm{ext}}$) below. It is essential that the value of such
penetration depth $r_{1}^{\mathrm{ext}}$\ itself does not appear explicitly
in the expression for the free energy of the system with the intruded
vortex, which is again accounted for by means of $y_{0}.$ 
It becomes of the order of $%
\ L$\ for weak enough currents and of the order of the effective coherence
length $\xi _{GL}(J)$\ (see Eq. (\ref{y0})) for the strong currents (see Fig. %
\ref{order}).

\section{Generalities}

In order to calculate the value of the activation energy $\delta F$ for the
current-biased NSS we start with the free energy functional $\mathcal{F}_{s}$%
\ including both GL and the current-field interaction terms (see Ref. \cite%
{G59}): 
\begin{align}
\mathcal{F}_{s}& =\nu \int d^{3}\mathbf{r}\left[ -\tau |\Delta \left( 
\mathbf{r}\right) |^{2}+\frac{\pi \mathcal{D}}{8T}|\partial _{-}\Delta
\left( \mathbf{r}\right) |^{2}\right.  \label{GLfull} \\
& \left. +\frac{7\zeta \left( 3\right) }{16\pi ^{2}T^{2}}|\Delta \left( 
\mathbf{r}\right) |^{4}\right] +\frac{1}{c}\int d^{3}\mathbf{r}\left( 
\mathbf{A}\left( \mathbf{r}\right) -\frac{c}{2e}\mathbf{\nabla }\varphi
\right) \cdot \mathbf{j}_{\infty }.  \notag
\end{align}%
Here $\Delta \left( \mathbf{r}\right) $ is the order parameter, $\mathbf{A}%
\left( \mathbf{r}\right) $ is vector-potential, $\nu =mp_{F}/\left( 2\pi
^{2}\right) $\ is the density of states ($p_{F}$\ is the electron Fermi
momentum), $\partial _{-}=\partial /\partial \mathbf{r-}2ie\mathbf{A/}c$, $%
\zeta \left( x\right) $\ is the Riemann zeta-function, $j_{\infty }$ $=J/S,$ 
$S$ is the cross-section of the stripe, $c$ is the speed of light, $\varphi $
is the phase of the order parameter. We use the system of units where $k_{%
\mathrm{B}}=1$ and $\hbar =1.$This functional allows to write down the
equations both for order parameter and vector-potential coordinate
dependencies.

\begin{figure}[th]
\includegraphics[width=1.01\columnwidth]{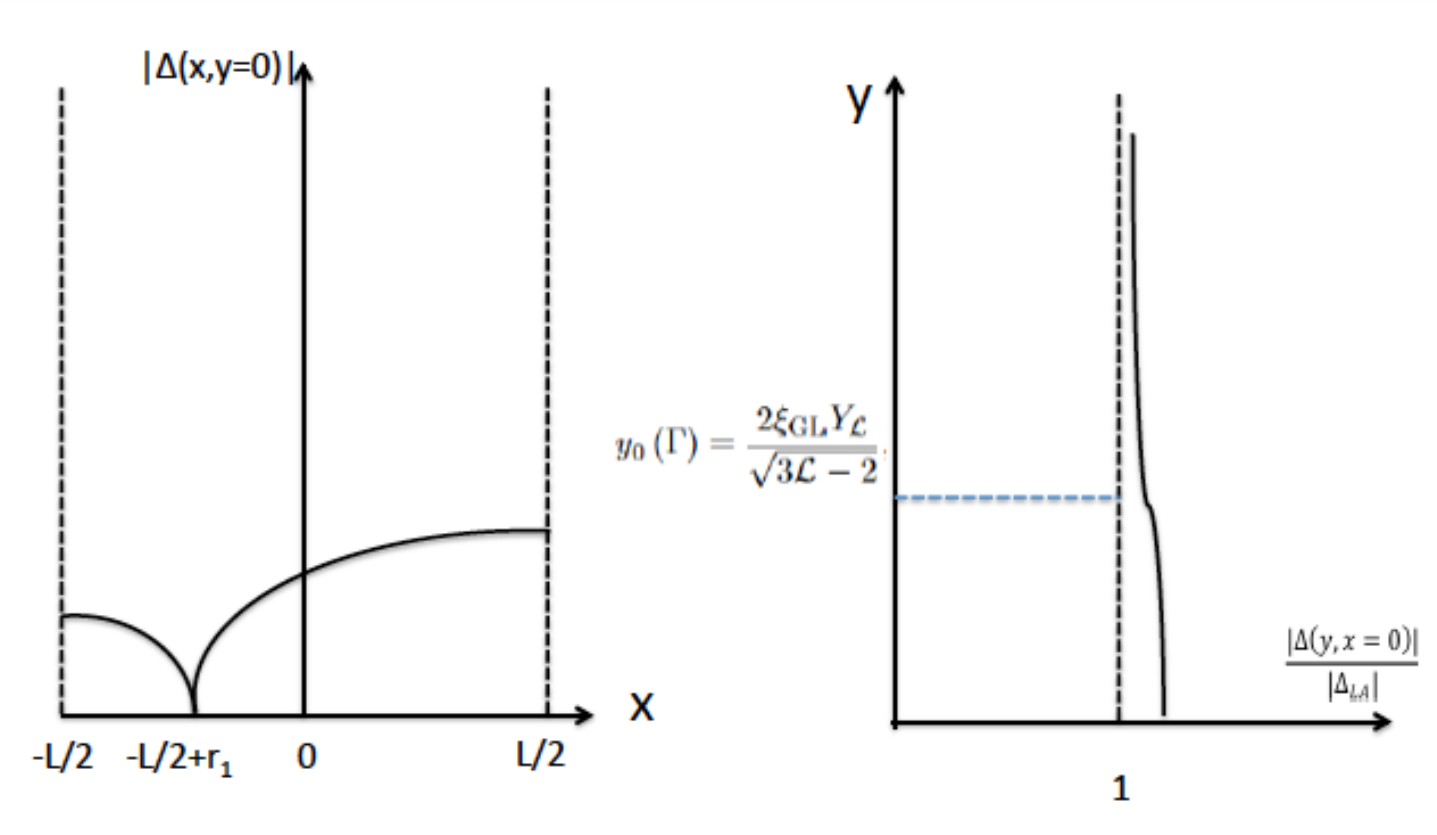}
\caption{Distribution of the modulus of the order parameter in the strip
with flowing current disturbed in the presence of a vortex: a) transversal
distribution at $y=0$; b) longitudinal distribution at $x=0$.}
\label{order}
\end{figure}

\begin{figure}[th]
\includegraphics[width=0.75\columnwidth]{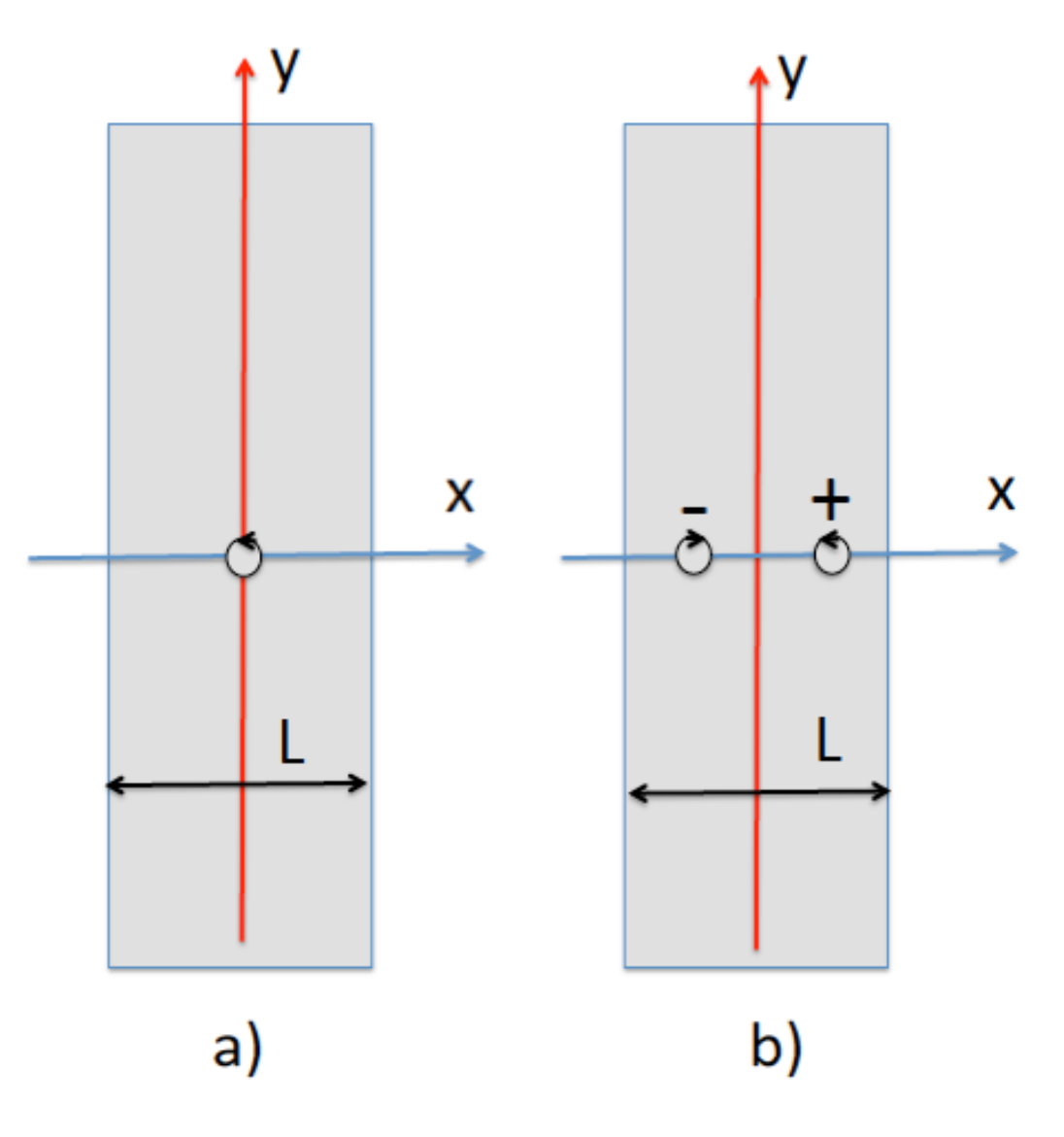}
\caption{Positions of the zeros of the saddle point solutions with one (a),
two (b), etc, vortices. Here by means of $\{+,-\}$ are denoted the
vorticities (phase factor changes by $\pm 2\protect\pi $ for
anticlock/clockwise circulation of the order parameter zeros).}
\label{stripe1}
\end{figure}

\emph{Close to zero current value.} Let us start with the simplest case of
zero current, $j_{\infty }=0.$ In this case an infinite number of 
saddle point solutions exist. If the saddle point solution has only one
zero corresponding to a single vortex state, symmetry considerations
show that the center of this vortex is located at the central line of the
strip (see Fig. \ref{stripe1}). Choosing the latter as the center of
coordinates one can find that the phase and modulus of the order parameter
are determined as:%
\begin{equation}
\left[ \frac{\partial \varphi \left( x,y\right) }{\partial \mathbf{r}}\right]
^{2}=\frac{\pi ^{2}}{L^{2}}\frac{1}{\sin ^{2}\left( \pi x/L\right) +\sinh
^{2}\left( \pi y/L\right) },  \label{phase}
\end{equation}%
\begin{equation}
\left\vert \Delta (x,y,y_{0})\right\vert =\frac{\pi \Delta _{0}}{L\cosh 
\frac{\pi y}{L}}\frac{1}{\sqrt{\left( \partial \varphi /\partial \mathbf{r}%
\right) ^{2}}}\tanh \frac{\sqrt{y^{2}+y_{0}^{2}}}{2\xi _{\mathrm{GL}}},\quad
\label{mod}
\end{equation}%
where $L$ is the width of the strip and $y_{0}$ is the free parameter
discussed at the end of the introduction (see also
Fig. \ref{order}). The phase of the order parameter  $\varphi$ is the solution of
the two dimensional Laplace equation $\Delta \varphi =0$ with boundary
conditions $\left. \partial \varphi /\partial x\right\vert _{x=\pm L/2}=0.$
The expression (\ref{mod}), obtained by means of the
variational procedure, coincides with the corresponding solution of the GL
equation in the limit $|y|\gg L$.

Substitution of Eqs. (\ref{phase}) and (\ref{mod}) to Eq. (\ref{GLfull})
gives the value of the activation energy versus $y_{0}:$%
\begin{widetext}
\begin{equation}
\delta F^{\left(  1\right)  }\left(  y_{0}\right)  \!=\!4\nu\Delta_{0}^{2}\tau
S\xi_{\mathrm{GL}}\left\{  \frac{2}{3}\!-\!\frac{\pi y_{0}}{8\xi_{\mathrm{GL}%
}}+\!\frac{L}{4\pi\xi_{\mathrm{GL}}}\left[  \frac{8\pi^{2}\left(
y_{0}/L\right)  ^{2}+4\zeta\left(  2\right)  -1}{6}\!\!+\!\frac{\pi^{2}%
y_{0}^{2}}{2L^{2}}I_{0}\left(  \!\frac{\pi y_{0}}{L}\right)  \right]
\right\}  ,\label{4a}
\end{equation}
\end{widetext}with 
\begin{equation*}
I_{0}\left( \!\alpha \right) =\int_{0}^{\infty }\frac{dx}{\cosh ^{2}x}\frac{1%
}{x^{2}+\alpha ^{2}}.
\end{equation*}%
The details of derivation of Eq. (\ref{4a}) are given in Appendix A. 

Minimization of Eq. (\ref{4a}) over $y_{0}$ gives the value $y_{0}=0.47L/\pi
;$ the corresponding value for the one-vortex configuration activation energy is
\begin{equation}
\delta F^{\left( 1\right) }=4\nu \Delta _{0}^{2}\tau S\xi _{\mathrm{GL}%
}\left( \frac{2}{3}+0.058\frac{L}{\xi _{\mathrm{GL}}}\right) .
\label{active1a}
\end{equation}%
An analogous consideration of the two vortices configuration gives for $%
\delta F^{\left( 2\right) }$ an answer similar to Eq. (\ref{active1a}) with
the  second term in brackets being twice smaller (see Appendix A). Further
increase of the  number of zeros in the order parameter results in the decrease of the
second term in $\delta F^{\left( n\right) }$ by the factor $n$ with respect to $%
\delta F^{\left( 1\right) }$. In the limit $n\rightarrow \infty $ the latter
reaches the value 
\begin{equation}
\delta F^{\left( \infty \right) }=8\nu \Delta _{0}^{2}\tau S\xi _{\mathrm{GL}%
}/3  \label{LA1}
\end{equation}%
first obtained in Ref. \cite{LA67} in the frameworks of the one-dimensional
model.

The flow of any finite current through the strip results in the finiteness
of the number of the saddle point solutions. This number rapidly decreases
with the current growth and already at so small current as $J_{\mathrm{c1}%
}=\eta\left( L/\xi_{GL}\right) J_{\mathrm{c}}$ the only saddle point solution
with one vortex remains. At higher currents the saddle point solutions do
not exist more, the critical points appear instead of them.

One can see that at zero current, the solution of the GL equation found in Ref. 
\cite{LA67} actually is the limiting one for the multiple-vortex solutions
obtained above. As a result we can state that the point $J=0$ is a singular
point in the current dependence of the activation energy. \ Therefore the
dependence of $\delta F^{\left( \infty \right) }(J)$, obtained in Ref. \cite%
{LA67} for small currents as the linear, in fact turns out to be substantially
more complicated. 
It is worth to mention that the multiplicity
of saddle point solutions in the domain of weak currents results in an increase
of the possibilities of the phase-slip events, i.e., to a noticeable
increase of the pre-exponential factor.

\section{``Weak'' currents}

Now we consider the most simple one-vortex state in the region of weak currents 
$J\leq J_{c1}.$ The vortex, corresponding to the saddle point solution, now
slightly shifts with respect to the  central axis of the strip. Denoting
the distance between the axis \ ($x=0$) and vortex center as $\delta $ ( $%
\delta =L/2-r_{1}$)$,$ we look for the solution of the Ginzburg-Landau
equation in the form%
\begin{equation}
\left\vert \Delta \right\vert =\Delta _{0}\left( \Gamma \right)
Z^{1/2}\left( \sqrt{y^{2}+y_{0}^{2}}\right) \Phi \left( x,y,\delta \right) ,
\label{modul}
\end{equation}%
with the functions%
\begin{widetext}
\begin{equation}
\Phi^{2}\left(  x,y,\delta\right)  =\frac{\sinh^{2}\left(  \pi y/L\right)
+\sin^{2}\left(  \pi x/L\right)  +\sin^{2}\left(  \pi\delta/L\right)
-2\cosh\left(  \pi y/L\right)  \sin\left(  \pi x/L\right)  \sin\left(
\pi\delta/L\right)  }{\cosh^{2}\left(  \pi y/L\right)  }\label{phi1}%
\end{equation}
\end{widetext}and 
\begin{equation}
Z\left( y,\Gamma \right) =1-\frac{3\mathcal{L}-2}{\mathcal{L}}\left[ \cosh
\left( \frac{y\sqrt{3\mathcal{L}-2}}{2\xi _{\mathrm{GL}}}\right) \right]
^{-2}.  \label{z}
\end{equation}%
Here $\Gamma =J/J_{c},$ while $\Delta _{0}\left( \Gamma \right) $ is the
order parameter of the homogeneous ground state of the NSS carrying on the
current $J$, i.e. the asymptotic form of our $\Delta \left( x,y,\Gamma
,y_{0,}\delta \right) $ far from the vortex, at $y\rightarrow \pm \infty $.
The latter can be related to the BCS value $\Delta _{00}{}\left( \tau
\right) =\left[ 8\pi ^{2}T^{2}\tau /\left( 7\zeta \left( 3\right) \right) %
\right] ^{1/2}$ of the order parameter in the absence of current by means of
the relation 
\begin{equation}
\Delta _{0}^{2}{}\left( \Gamma \right) =\Delta _{00}^{2}\mathcal{L}\left(
\Gamma \right) .  \label{pm}
\end{equation}%
The choice of the two former multipliers in the anzatz (\ref{modul}) is
based on the Langer-Ambegaokar solution of the GL equation for current
biased one-dimensional channel distorted by the vortex presence (and
accounting for its evenness in $y$). The latter multiplier (see Eq. (\ref%
{phi1})) accounts for the appearing in the case under consideration
asymmetry of the order parameter dependence on the transversal coordinate
and in the case of $\delta =0$ it leads to the coincidence of Eq. (\ref%
{modul}) and Eqs. (\ref{mod})-(\ref{phase}). Substitution of the anzatz (\ref%
{modul}), (\ref{phi1}), and (\ref{z}) to the GL equations gives the explicit
value for $\mathcal{L}\left( \Gamma \right) :$ 
\begin{equation}
\mathcal{L}\left( \Gamma \right) =\frac{1}{3}+\frac{2}{3}\sin \left[ \frac{%
\pi }{6}+\frac{2}{3}\arcsin \sqrt{1-\Gamma ^{2}}\right] .  \label{Lemp}
\end{equation}%
The quantity $y_{0}\sim L,$ \ hence for $\left\vert y\right\vert \gg L$ the
modulus of the order parameter $\left\vert \Delta \right\vert $ (see Eq. (%
\ref{modul})) should obey the exact GL equations with the corresponding
boundary conditions at $x=\pm L/2.$ The function $\Phi $ is related to the
order parameter phase $\varphi $ which satisfies the equation%
\begin{equation}
\left( \frac{\partial \varphi \left( x,y\right) }{\partial \mathbf{r}}%
\right) ^{2}=\left( \frac{\pi }{L}\right) ^{2}\frac{\cos ^{2}\frac{\pi
\delta }{L}}{\cosh ^{2}\left( \frac{\pi y}{L}\right) \Phi ^{2}\left(
x,y,\delta \right) }.  \label{phase_weak}
\end{equation}

Substitution of the Eqs. (\ref{modul}), (\ref{phi1}), and (\ref{phase_weak})
into Eq. (\ref{GLfull}) leads to the expression for free energy 
\begin{widetext}%
\begin{align}
\delta F_{\delta}\left(  y_{0}\right)   &  =4\nu\Delta_{0}^{2}\tau
S\xi_{\mathrm{GL}}\left\{  \frac{2}{3}-\frac{\pi y_{0}}{8\xi_{\mathrm{GL}%
}}-\frac{16\pi\xi_{\mathrm{GL}}\Gamma^{2}}{27L}\left(  \frac{\pi}{4a}+\frac
{L}{8y_{0}}+I_{1}\left(a\right)  \right)  + \frac{L}{4\pi\xi
_{\mathrm{GL}}}\left[  \frac{4a^{2}}{3}+\frac{2}{3}\zeta\left(  2\right)
-\frac{1}{6}+\frac{a^{2}}{2}I_{0}\left(  \!a\right)  \right]  \right.
\nonumber\\
&  \left.  +\frac{\pi\Gamma}{3}\sqrt{\frac{2}{3}}\sin\left(  \pi
\delta/L\right)  -\frac{L}{4\pi\xi_{\mathrm{GL}}}\left[  \frac{2a^{2}}%
{3}+\frac{1}{3}\zeta\left(  2\right)  + a^{2}I_{0}\left( a\right)
\right]  \sin^{2}\left(  \pi\delta/L\right)  \right\}  ,\label{long}%
\end{align}
\end{widetext}where%
\begin{equation*}
a^{2}=\frac{\pi^{2}y_{0}^{2}}{L^{2}}+\frac{32\pi^{2}\Gamma^{2}}{27L^{2}}\xi_{%
\mathrm{GL}}^{2}
\end{equation*}
and%
\begin{equation*}
I_{1}\left( \!a\right) =\int_{0}^{\infty}\frac{dx}{\cosh x}\frac{1}{%
x^{2}+4a^{2}}.
\end{equation*}
The details of transition from Eq. (\ref{GLfull}) to Eq. (\ref{long}) are
presented in the Appendix A.

Let us recall, that the quantities $\left\{ y_{0},\delta \right\} $\ still
remain indefinite: their values one can determine from the conditions of the
GL\ functional $\delta F_{\delta }\left( y_{0}\right) $ extremum:%
\begin{equation}
\frac{\partial \delta F_{\delta }\left( y_{0}\right) }{\partial y_{0}}=0,\;
\label{optim1}
\end{equation}%
\begin{equation}
\frac{\partial \delta F_{\delta }\left( y_{0}\right) }{\partial \delta }=0
\label{optim2}
\end{equation}%
In result of solution of Eq. (\ref{optim2}) \ the value $\delta $ can be
presented as the function of $\Gamma :$%
\begin{equation}
\sin \frac{\pi \delta }{L}=\left( \frac{2}{3}\right) ^{3/2}\frac{\pi ^{2}\xi
_{\mathrm{GL}}\Gamma }{L\left[ 2a^{2}/3+\zeta \left( 2\right)
/3+a^{2}I_{0}\left( a\right) \right] }.  \label{optim3}
\end{equation}%
What concerns the value $y_{0}$ it is determined by Eqs. (\ref{optim1}) and (%
\ref{optim3}). Corresponding equation is very cumbersome and we do not
present it here. Important, that it has the solution only in the very narrow
currents interval 
\begin{equation*}
\Gamma ^{2}\leq \Gamma _{c1}^{2}=0.009\frac{L^{2}}{\xi _{\mathrm{GL}}^{2}}.
\end{equation*}%
Finally, the value of the free energy $\delta F_{\delta }$ in the critical
point $J=J_{c1}$ is 
\begin{equation}
\delta F_{\delta }\left( \tau ,J_{c1}\right) =\!4\nu \Delta _{0}^{2}\left(
\Gamma \right) \tau S\xi _{\mathrm{GL}}\left( \frac{2}{3}+0.054\frac{L}{\xi
_{\mathrm{GL}}}\right) .  \label{fw}
\end{equation}%
Comparing the Eqs. (\ref{fw}) and (\ref{active1a}) one can see that the
state with $J=J_{c1}$, when the only saddle point solution with one vortex
remains, energetically differs from that one with $J=0$ by very small
quantity $0.004\left( \frac{L}{\xi _{\mathrm{GL}}}\right) 4\nu \Delta
_{0}^{2}\left( \Gamma \right) \tau S\xi _{\mathrm{GL}}$.

\section{``Strong'' currents}

Let us pass to discussion of the mechanism of energy dissipation in the wide
range of currents $J_{c1}\ll J<J_{c},$ when the GL equations do not have
more any saddle point solution. Let us suppose that through the edge of the
strip penetrates a single vortex and assume that its center is located at
some small distance $r_{1}(r_{1}\ll L)$ from the edge, i.e. the vortex
center coordinates are: $\left( -L/2+r_{1},0\right) $. Our goal is to obtain
the maximal possible value of the \textquotedblleft penetration
length\textquotedblright\ $r_{1}$ at which the requirement of existence of
the conditional extremum of the functional (\ref{GLfull}) is still
satisfied. In order to do this we look for the phase and the modulus of the
order parameter in the form containing three free parameters: $%
r_{1},y_{0},\gamma $ 
\begin{equation}
\left( \frac{\partial \varphi \left( x,y,r_{1}\right) }{\partial \mathbf{r}}%
\right) ^{2}=\left( \frac{\pi }{L}\right) ^{2}\frac{\sin ^{2}\frac{\pi r_{1}%
}{L}}{\cosh ^{4}\left( \frac{\pi y}{2L}\right) }\left[ Q(x,y)\right] ^{-1},
\label{phl}
\end{equation}%
and%
\begin{equation}
\left\vert \Delta \left( x,y,r_{1},y_{0},\gamma \right) \right\vert
\!=\!\Delta _{0}\left( \Gamma \right) \frac{\ln \left[ 1\!+\!\frac{\gamma
L^{2}}{r_{1}^{2}}Q^{1/2}(x,y)\right] }{\ln \left( \frac{2\gamma L^{2}}{%
r_{1}^{2}}+1\right) }\!Z^{\frac{1}{2}}\!\left( \Gamma ,\!y\!+\!y_{0}\right)
\label{dl}
\end{equation}%
The function 
\begin{align}
Q(x,y,r_{1})& =\left\{ 4\sinh ^{2}\left( \frac{\pi y}{2L}\right) \left[
\cosh ^{2}\frac{\pi y}{2L}+\cos \frac{\pi r_{1}}{L}\sin \frac{\pi x}{L}%
\right] \right.  \notag \\
& \left. +\left[ \sin \frac{\pi x}{L}+\cos \frac{\pi r_{1}}{L}\right]
^{2}\right\} \left[ \cosh \left( \frac{\pi y}{2L}\right) \right] ^{-4},
\label{qxy}
\end{align}%
is the result of direct calculation of the phase $\varphi $ in the
one-vortex state of the strip (see Eq. (\ref{phl})). The function $%
\left\vert \Delta \left( x,y\right) \right\vert $ approaches to the solution
of the GL equation in the range $\left\vert y\right\vert \gg L$. Both Eqs. (%
\ref{phl}) and (\ref{dl}) satisfy the boundary conditions for the order
parameter and its derivatives at the edge of the stripe and at infinity.
What concerns the variational parameter $\gamma $\ it can be found from the
condition%
\begin{equation}
\frac{\partial \delta F\left( r_{1},y_{0},\gamma \right) }{\partial \gamma }%
=0.  \label{gammaex}
\end{equation}%
It determines the shape of the order parameter and, correspondingly, the
contribution to the free energy from the domain close to the vortex $%
|y|\lesssim L$. Its introduction allows to improve the variational
approximation in this region. Corresponding expression turns to be of the
order of 1 and does not appear explicitly in the final expression for the
free energy, it is why we do not present it here.

The current conservation law leads to the next expression for the essential
part of the vector potential $\mathbf{A}$%
\begin{equation}
A_{y}^{\left( 0\right) }\left( y\right) =\frac{A_{\infty }\Delta
_{0}^{2}{}\left( \Gamma \right) }{\left\langle |\Delta \left( x,y\right)
|^{2}\right\rangle _{x}{}},  \label{vector}
\end{equation}%
where $\left\langle ....\right\rangle _{x}$ means the averaging over the
transverse coordinate. The value of the vector potential at infinity $%
A_{\infty }$ is determined by the current density:%
\begin{equation}
\frac{J}{S}=-\frac{\pi \nu e^{2}\mathcal{D}}{T}\Delta _{0}^{2}{}\left(
\Gamma \right) A_{\infty }.
\end{equation}%
Replacing the expressions (\ref{phl}) , (\ref{dl}) , and (\ref{vector}) to
the GL\ functional (\ref{GLfull}) and calculating the integral one can find
the value of excess free energy $\delta F_{s}$ of the strip with the
penetrated vortex with respect to its the ground state with the fixed
current. The requirement of existence of the conditional extremum determines
the value $y_{0}:$%

\begin{equation}
y_{0}\left( \Gamma \right) =\frac{2\xi _{\mathrm{GL}}Y_{\mathcal{L}}}{\sqrt{%
3 \mathcal{L}-2}},  \label{y0}
\end{equation}

\begin{equation*}
\tanh Y_{\mathcal{L}}\left( \Gamma \right) =2\left[ \frac{\left( 1-\mathcal{L%
}\right) }{4-3\mathcal{L}+\sqrt{\mathcal{L}\left( 16-15\mathcal{L}\right) }}%
\right] ^{1/2}.\quad
\end{equation*}%
The main steps of this calculus are presented in the Appendix B. The
obtained results allow us to write down the expression for the activation
energy in the whole region of \textquotedblleft strong
currents\textquotedblright\ \ $J_{c1}\ll J<J_{c}$ :

\begin{widetext}%
\begin{align}
\delta F\left(  \tau,J\right)   &  =4\nu\Delta_{0}^{2}\left(  \Gamma\right)
\tau S\xi_{\mathrm{GL}}\sqrt{\left(  3\mathcal{L}-2\right)  }\left\{
\frac{1-\tanh Y_{\mathcal{L}}}{6\mathcal{L}}\left[  4+\left(  3\mathcal{L}%
-2\right)  \tanh Y_{\mathcal{L}}\left(  1+\tanh Y_{\mathcal{L}}\right)
\right]  \right.  \nonumber\\
&  -\left.  \sqrt{\frac{2\left(  1-\mathcal{L}\right)  }{3\mathcal{L}-2}%
}\left[  \arctan\sqrt{\frac{3\mathcal{L}-2}{2\left(  1-\mathcal{L}\right)  }%
}-\arctan\left(  \sqrt{\frac{3\mathcal{L}-2}{2\left(  1-\mathcal{L}\right)  }%
}\tanh Y_{\mathcal{L}}\right)  \right]  \right\}  .\label{dl2}%
\end{align}
\end{widetext}It is seen that the difference between Eq. (\ref{dl2}) and the
expression for the activation energy of the one-dimensional superconducting
channel carrying current $J$ (the main result of Ref. \cite{LA67}) consists
of the contribution occurring due to the nonzero value of the parameter $Y_{%
\mathcal{L}},$ i.e. due to existence of the conditional extremum of the free
energy functional at the distance $y_{0}\neq 0$. Let us stress that the
activation energy $\delta F$ depends on the geometry of a sample, which here
is assumed as a strip. The increase of the energy barrier in the Arrenius
law with respect to the result of Ref. \cite{LA67} is related to\ the
necessity of the vortex penetration in a sample at the moment of the phase
slip event.

The expression for activation energy $\delta F^{\left( LA\right) }\left(
\tau ,J\right) $ of the one-dimensional superconducting channel found in
Ref. \cite{LA67} can be easily reproduced from Eq. (\ref{dl2}) just putting $%
Y_{\mathcal{L}}=0$ (what follows from Eq. (\ref{y0})). \ One can compare the
result of our careful consideration of the vortex penetration mechanisms (%
\ref{dl2}) with the latter: 
\begin{widetext}%
\begin{equation}
\frac{\delta F\left(  \tau,J\right)  -\delta F^{\left(  LA\right)  }\left(
\tau,J\right)  }{\delta F^{\left(  LA\right)  }\left(  \tau,J\right)  }%
=\frac{\left[  \frac{3\mathcal{L}-2}{6\mathcal{L}}\frac{\tanh Y_{\mathcal{L}}%
}{\cosh^{2}Y_{\mathcal{L}}}-2\frac{\tanh Y_{\mathcal{L}}}{3\mathcal{L}}%
+\sqrt{\frac{2\left(  1-\mathcal{L}\right)  }{3\mathcal{L}-2}}\arctan\left(
\sqrt{\frac{3\mathcal{L}-2}{2\left(  1-\mathcal{L}\right)  }}\tanh
Y_{\mathcal{L}}\right)  \right]  }{\left[  \frac{2}{3\mathcal{L}}-\sqrt
{\frac{2\left(  1-\mathcal{L}\right)  }{3\mathcal{L}-2}}\arctan\sqrt
{\frac{3\mathcal{L}-2}{2\left(  1-\mathcal{L}\right)  }}\right]
}.\label{difen}%
\end{equation}
\end{widetext}.

\begin{figure}[bh]
\includegraphics[width=\columnwidth]{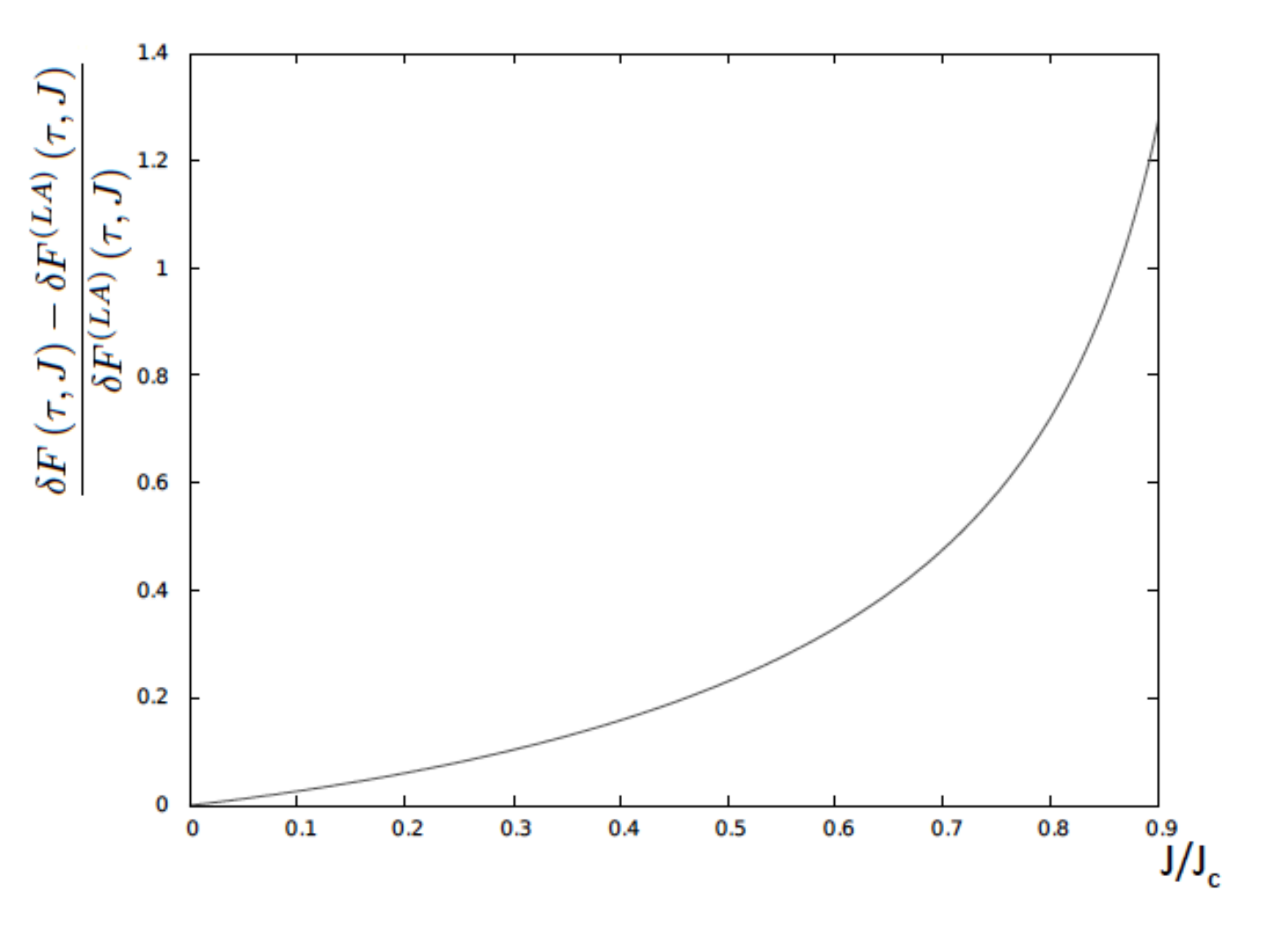}
\caption{Excess activation energy related to the account for the true
mechanism of the vortices penetration in the strip as the function of
flowing current.}
\label{diff}
\end{figure}

For the currents larger than $J_{c1}$\ but still much smaller than $J_{c}$\
the saddle point solutions of GL equations considered in Ref. \cite{LA67} do
not exist more. Nevertheless one can see that the difference between the
free energy of the conditional extremum and that one calculated by Langer
and Ambegaokar (see Eq. (\ref{difen})) turns to be proportional only to $%
\left( J/J_{c}\right) ^{2},$\ i.e. the result of Ref. \cite{LA67} remains
valid. The situation considerably changes when the current approaches its
critical value, $J\rightarrow J_{c}$($\Gamma \rightarrow 1$). Here $\tanh Y_{%
\mathcal{L}}\rightarrow 1/\sqrt{3};\cosh ^{2}Y_{\mathcal{L}}\rightarrow
3/2;1+2^{-1}\cosh ^{-2}Y_{\mathcal{L}}-L\tanh ^{2}Y_{\mathcal{L}}/\left[
2\left( 1-\mathcal{L}\right) \right] \rightarrow 1;3L-2\rightarrow 2^{3/2}%
\sqrt{1-\Gamma }/\sqrt{3}$, and

\begin{equation}
\delta F^{\left( LA\right) }\left( \tau ,\Gamma \rightarrow 1\right) =\frac{4%
}{15}\sqrt{\frac{2}{\sqrt{3}}}\nu \tau \Delta _{00}^{2}\left( \tau \right)
S\xi _{\mathrm{GL}}\left( 1-\Gamma ^{2}\right) ^{5/4},  \label{LAclose}
\end{equation}%
while%
\begin{equation}
\delta F\left( \tau ,\Gamma \rightarrow 1\right) =\frac{2^{3/2}}{3^{9/4}}\nu
\tau \Delta _{00}^{2}\left( \tau \right) S\xi _{\mathrm{GL}}\left( 1-\Gamma
^{2}\right) ^{3/4}.  \label{OVclose}
\end{equation}%
The relative difference of the free energies Eq. (\ref{difen}) in this case\
diverges:%
\begin{equation}
\left. \frac{\delta F\left( \tau ,J\right) -\delta F^{\left( LA\right)
}\left( \tau ,J\right) }{\delta F^{\left( LA\right) }\left( \tau ,J\right) }%
\right\vert _{J\rightarrow J_{c}}=\frac{5J_{c}}{6\sqrt{J_{c}^{2}-J^{2}}},
\label{normal}
\end{equation}%
\ i.e. the height of the barrier in Arrenius law turns out to be
parametrically larger than predicted in Ref. \cite{LA67}. The behavior of
Eq. (\ref{difen}) \ as the function of $\Gamma $\ in the interval $\left[
0,0.9\right] $\ is presented in Fig. \ref{diff}. 

\section{Pre-exponential factor}

In order to obtain the exact value of the pre-exponential factor $\Omega$
for phase slip events one should have in possession the expression for the
effective action of superconducting strip containing vortices. In the Ref. 
\cite{LO84} was proposed a general procedure which, in the regime of thermal
fluctuations, is reduced to solution of the spectral problem for linear
operator corresponding to the action at its saddle point. The difficulty of
the problem under consideration consists in the fact that nor microscopic
action operator is known nor saddle point (for currents $J_{c1}<J)$ exists.
Nevertheless, the knowledge of action would allow one to get the precise
value of $\ \Omega$ at least for weak currents $J<J_{c1},$ while for strong
currents one could believe that change of the saddle point to a singular
point would not strongly effect on the value of pre-exponential factor. In
light of the above said the evaluations of $\Omega$ in both papers \cite%
{LA67,MH70} seem doubtful: use of the time-dependent GL equation below $T_{%
\mathrm{c}}$ as today is well known cannot be justified, unless in gapless regime\cite{GE68}.

The main contribution to the average time between two subsequent phase-slip
events is related to the existence of the ``zero-mode'' (see Ref. \cite%
{MH70,LO84}). In the case under consideration the size of the vortex is
determined by the transversal size $L$ of the strip. The vortices which
slip on the distances larger than $L$ can be considered as independent. It
is why the main factor determining the pre-exponential one is the ratio of
the transversal size $L$ to the strip length $\mathfrak{L}$: $L/\mathfrak{L.%
}$ Another coefficient which forms the pre-exponential factor is the
characteristic ``crossing time'' $\Delta t_{\mathrm{cross}}=L/v_{\mathrm{%
cross}}$ of the strip by the vortex, moving with the velocity \cite{LO86}:%
\begin{equation}
v_{\mathrm{cross}}=\frac{cJ\sqrt{\tau}}{4SH_{c2}\sigma_{n}},  \label{vcross}
\end{equation}
where $\sigma_{n}$ is the conductivity of the strip in its normal phase.
Finally, accounting for Arrenius factor, one finds the characteristic time $%
\Delta t$ between two phase slipping events

\begin{equation}
\Delta t=\frac{L}{\mathfrak{L}}\left( \frac{L}{v_{\mathrm{cross}}}\right)
\exp \left( \frac{\delta F}{k_{\mathrm{B}}T}\right) .  \label{deltat}
\end{equation}%
The average voltage $V$ at the strip is related to the average time
interval $\Delta t$ between the voltage jumps by the Josephson relation\cite%
{J62}: $V=\pi \hbar /\left( e\Delta t\right) .$ Corresponding resistance of
the strip is%
\begin{equation}
\frac{R}{R_{0}}=\frac{\pi \hbar c\sqrt{\tau }}{4eH_{c2}L^{2}}\exp \left[ -%
\frac{\delta F\left( \tau ,\Gamma \right) }{k_{\mathrm{B}}T}\right] ,
\label{arrenius}
\end{equation}%
where $R_{0}=\mathfrak{L}/\left( \sigma _{n}S\right) $ is the normal
resistance of the strip. It is necessary to mention that the used above
approximation of the independent phase slips is valid only (according to Eq.
(\ref{deltat}) ) when $\delta F\gtrsim k_{\mathrm{B}}T.$

\section{Conclusions}

We have demonstrated that considering only the longitudinal spatial dependence of the order parameter
in narrow superconducting strips carrying finite current (see Ref. \cite{LA67}) is not sufficient to describe the
properties of its resistive state correctly. Taking the transversal coordinate into account when calculating the saddle point solutions of the GL equation turns out to be essential.
Namely, the value of activation energy in the Arrenius law for the resistance of a narrow superconducting channel differs
\textit{already for relatively weak currents} from the value obtained by simply using the
difference of the free energy of such a saddle point and the ground state energy.
The mechanism for phase-slip events turns out to be much more
sophisticated then the one described in Ref.~\cite{LA67}.

Already at weak currents ($J<J_{\mathrm{c1}}$) a sequence of the saddle
points appears, which is characterized by the number $n$\ of zeros of the order parameter along
the transverse coordinate. The energy of such state equals to that
one found by Langer and Ambegaokar \cite{LA67} only in the limit $%
n\rightarrow \infty ,$\ when the system carries no current. 
One could say that the state of the strip in the current-free case is singular. 
The number $n$ of saddle points rapidly decreases with the growth of the current. 
It reaches $n=1$ when the current has the value $J_{\mathrm{c1}%
}=0.0312\left( L/\xi \right) J_{\mathrm{c}}$: \ at this point only
a stationary state remains.

When $J>J_{\mathrm{c1}}$, stationary solutions of the GL equations
with fixed current and a vortex in the strip do not exist.
Instead one needs to look for a critical point, corresponding to the existence of
a specific conditional extremum of the GL functional. 
These conditions are: the current, $J$, is fixed and the distance between the
vortex center and the strip edge is maximal. The energy of such a state turns out to
be larger than the activation energy $\delta F^{\left( LA\right) }\left(
\tau ,J\right) $\ obtained in Ref. \cite{LA67}. The normalized difference (%
\ref{normal}) increases with growth of the current and when the latter
approaches to the critical value $J_{c}$\ the former diverges (see Fig. \ref%
{diff}).

Experimentally, the discrepancy between the theoretical prediction of
Ref. \cite{LA67} and the mechanism proposed here can be detected by analyzing the current dependence of the resistance close to $J_{\mathrm{c}%
}.$\ 
The predicted dependence of $\log R(I)$ with exponent $5/4$ in Ref. \cite{LA67}, should transform into
a weaker one with exponent $3/4$ in the region $J\rightarrow J_{c}$.
Some experimental papers indicated an unexpected decrease of the resistance in the regime of strong
currents (see Ref. \cite{T75,B12,B14}). This long standing enigma can potentially be resoled by the above
analysis. 
A recent numerical study narrow superconducting channels using the time-dependent GL equation in the strong current regime also indicated that the critical exponent of the activation energy is of the order of 0.7 for the widths $L\sim \xi _{GL}$
(see Ref. \cite{V12}).

\acknowledgements

The authors acknowledge financial support of the FP7-IRSES program, grant N
236947 \textquotedblleft SIMTECH\textquotedblright . A.V. was partially supported by
the U.S. Department of Energy, Office of Science, Office of Advanced
Scientific Computing Research and Materials Sciences and Engineering
Division, Scientific Discovery through Advanced Computing (SciDAC) program.
A.V. is grateful to I. Aronson, A. Bezryadin and A. Glatz for valuable
discussions.

\appendix

\section{}

In the process of derivation of Eq. (\ref{4a}) we used the following
integrals:%
\begin{equation}
\frac{1}{L}\int_{-L/2}^{L/2}\frac{dx}{\sin^{2}\frac{\pi x}{L}+\!\sinh^{2}%
\frac{\pi y}{L}}\!=\!\frac{2}{\sinh\left( \frac{2\pi y}{L}\right) },
\label{A1}
\end{equation}%
\begin{equation}
\frac{1}{L}\int_{-L/2}^{L/2}\frac{\sin^{2}\left( \pi x/L\right) dx}{\sin ^{2}%
\frac{\pi x}{L}+\!\sinh^{2}\frac{\pi y}{L}}=\frac{\exp\left( \frac{-\pi y}{L}%
\right) }{\cosh\frac{\pi y}{L}},  \label{A2}
\end{equation}%
\begin{equation*}
\frac{1}{L}\int_{-L/2}^{L/2}\frac{\sin^{2}\left( \pi x/L\right) dx}{\left[
\sin^{2}\frac{\pi x}{L}+\!\sinh^{2}\frac{\pi y}{L}\right] ^{2}}=\frac {1}{%
\sinh\left( \frac{2\pi y}{L}\right) \cosh^{2}\left( \frac{\pi y}{L}\right) },
\end{equation*}%
\begin{equation*}
\frac{1}{L}\int_{-L/2}^{L/2}\frac{\sin^{4}\left( \pi x/L\right) dx}{\left[
\sin^{2}\frac{\pi x}{L}+\!\sinh^{2}\frac{\pi y}{L}\right] ^{3}}=\frac{3}{8}%
\frac{1}{\sinh\left( \frac{\pi y}{L}\right) \cosh^{5}\left( \frac{\pi y}{L}%
\right) },
\end{equation*}
for $y>0.$

Using these relations in the Eqs . (\ref{phase}) and (\ref{mod}) one finds

\begin{widetext}
\begin{align*}
\left\langle \frac{\partial|\Delta\left(  x,y\right)  |^{2}}{\partial
\mathbf{r}}\right\rangle _{x} &  =\Delta_{0}^{2}\left\{  \left(  \frac{\pi
}{2L}\right)  ^{2}\tanh^{2}\frac{\sqrt{y^{2}+y_{0}^{2}}}{2\xi_{\mathrm{GL}}%
}\frac{1-\exp\left(  -4\pi y/L\right)  }{\sinh\frac{2\pi y}{L}\cosh^{2}%
\frac{\pi y}{L}}+\frac{y^{2}}{8\left(  y^{2}+y_{0}^{2}\right)  }\frac
{1+\tanh^{2}\left(  \pi y/L\right)  }{\xi_{\mathrm{GL}}^{2}\cosh^{4}%
\frac{\sqrt{y^{2}+y_{0}^{2}}}{2\xi_{\mathrm{GL}}}}\right.  \\
&  \left.  +\left(  \frac{\pi}{2L}\right)  ^{2}\tanh^{2}\left(  \frac
{\sqrt{y^{2}+y_{0}^{2}}}{2\xi_{\mathrm{GL}}}\right)  \frac{\tanh^{2}\left(
\pi y/L\right)  }{\sinh\left(  2\pi y/L\right)  \cosh^{2}\left(  \pi
y/L\right)  }\left[  3+4\exp\left(  -2\pi y/L\right)  +\exp\left(  -4\pi
y/L\right)  \right]  \right.  \\
&  \left.  +\frac{\pi}{2\xi_{\mathrm{GL}}L}\frac{y}{\sqrt{y^{2}+y_{0}^{2}}%
}\tanh\left(  \frac{\sqrt{y^{2}+y_{0}^{2}}}{2\xi_{\mathrm{GL}}}\right)
\frac{\tanh\left(  \pi y/L\right)  }{\cosh^{2}\left(  \pi y/L\right)
\cosh^{2}\left[  \sqrt{y^{2}+y_{0}^{2}}/\left(  2\xi_{\mathrm{GL}}\right)
\right]  }\right\}  ,
\end{align*}
\end{widetext}for $y>0$. \ Here we introduced the symbol of averaging over
the transverse coordinate:%
\begin{equation*}
\left\langle \left( ...\right) \right\rangle _{x}=\frac{1}{L}\int
_{-L/2}^{L/2}dx\left( ...\right) .
\end{equation*}

Next we present \ the explicit integrals of the type (\ref{A1}) and (\ref{A2}%
) over $y:$%
\begin{align*}
& \int_{0}^{\infty}\frac{\left( y^{2}\!+\!y_{0}^{2}\right) \tanh^{2}\frac{%
\pi y}{L} }{\sinh\frac{2\pi y}{L} \cosh^{2} \frac{\pi y}{L} }\left[
3\!+\!4\exp\left( \! -\!\frac{2\pi y}{L}\right) \!+\!\exp\left( \! -\!\frac{%
4\pi y}{L}\right) \right] dy \\
& =\frac{2L}{\pi}\left[ y_{0}^{2}\left( \frac{5}{3}-2\ln2\right) +\left( 
\frac{L}{\pi}\right) ^{2}\left( \frac{5}{6}\zeta\left( 2\right) -\frac {5}{4}%
\zeta\left( 3\right) \right) -\frac{1}{3}\right] .
\end{align*}%
\begin{equation*}
\int_{0}^{\infty}\frac{dy}{\cosh^{4}\frac{\sqrt{y^{2}+y_{0}^{2}}}{2\xi_{%
\mathrm{GL}}}}\frac{y^{2}}{y^{2}+y_{0}^{2}}=\frac{4}{3}\xi _{\mathrm{GL}}-%
\frac{\pi}{2}y_{0},
\end{equation*}
$\;y_{0}\ll\xi_{\mathrm{GL}}.$

In the two vortex state with zero current, instead of Eqs . (\ref{phase})
and (\ref{mod}) one finds 
\begin{equation*}
\left[ \frac{\partial\varphi\left( x,y\right) }{\partial\mathbf{r}}\right]
^{2}=\frac{4\pi^{2}}{L^{2}}\frac{1}{\cos^{2}\frac{2\pi x}{L}+\sinh^{2}\frac{%
2\pi y}{L}},
\end{equation*}%
\begin{equation*}
\left\vert \Delta\right\vert =\Delta_{0}\tanh\frac{\sqrt{y^{2}+y_{0}^{2}}}{%
2\xi_{\mathrm{GL}}}\phi,\quad
\end{equation*}%
\begin{equation*}
\phi=\frac{1}{\cosh\frac{2\pi y}{L}}\left[ \cos^{2}\frac{2\pi x}{L}+\sinh
^{2}\frac{2\pi y}{L}\right] ^{1/2}
\end{equation*}
All following considerations are similar to those in a single vortex state.
In the domain of weak currents the current conservation law gives in the
main approximation the expression for the vector potential%
\begin{equation}
A\!=\!\left( 0,\frac{A_{\infty}\Delta_{0}^{2}\left( \Gamma\right) }{\left\langle
\left\vert \Delta\right\vert ^{2}\right\rangle _{x}}\!,0\right) ,\;\Gamma\!=\frac{J}{J_c}=\!-3%
\sqrt{6}\left\vert e\right\vert A_{\infty}\xi_{\mathrm{GL}}.  \label{A7}
\end{equation}
Next important moment is the calculus of the integrals of the type $%
\left\langle \Phi^{-2}\left( x,y\right) \right\rangle _{x}.$ One finds 
\begin{widetext}
\begin{align}
&  I_{b}=\frac{1}{L}\int_{-L/2}^{L/2}\frac{dx}{\sinh^{2}\frac{\pi y}{L}%
+\sin^{2}\frac{\pi x}{L}+\sin^{2}\frac{\pi\delta}{L}-2\sin\frac{\pi x}{L}%
\sin\frac{\pi\delta}{L}\cosh\frac{\pi y}{L}} \nonumber\\
&  \!=\!-\!4\frac{z_{1}z_{2}\left[  \left(  z_{3}\!+\!z_{4}\right)
\!-\!2\cosh\frac{2\pi y}{L}\!-\!4\sin^{2}\frac{\pi\delta}{L}\right]
\!+\!\left(  z_{1}\!+\!z_{2}\right)  \left[  \left(  1\!-\!z_{3}z_{4}\right)
\!+\!4z_{3}z_{4}\sin^{2}\frac{\pi\delta}{L}\!+\!2z_{3}z_{4}\cosh\frac{2\pi
y}{L}\!-\!\left(  z_{3}\!+\!z_{4}\right)  \right]  }{\left(  z_{1}%
\!-\!z_{3}\right)  \left(  z_{2}-z_{3}\right)  \left(  z_{1}-z_{4}\right)
\left(  z_{2}-z_{4}\right)  },\label{int14}
\end{align}
\end{widetext}
where%
\begin{equation*}
z_{1,2}=\exp\left( \frac{2\pi y}{L}\right) \left[ 1\pm2i\sin\frac{\pi \delta%
}{L}-2\sin^{2}\frac{\pi\delta}{L}\right] ,
\end{equation*}%
\begin{equation*}
z_{3,4}=\exp\left( -\frac{2\pi y}{L}\right) \left[ 1\pm2i\sin\frac {\pi\delta%
}{L}-2\sin^{2}\frac{\pi\delta}{L}\right] .
\end{equation*}
The direct and combersome integration of Eq. (\ref{int14}) \ results in 
\begin{equation}
I_{b}=\frac{2}{\sinh\frac{2\pi y}{L}}\left[ 1+\frac{4\sin^{2}\frac{\pi\delta 
}{L}\sinh^{2}\frac{\pi y}{L}}{\sinh^{2}\frac{2\pi y}{L}+4\sin^{2}\frac {%
\pi\delta}{L}}\right] .  \notag
\end{equation}
Now, using the Eqs. (\ref{A7})-(\ref{int14}), and the definition (\ref{z})
one can find the necessary values:

\begin{equation}
\left\langle \left[ \frac{\partial\varphi\left( x,y\right) }{\partial 
\mathbf{r}}\right] ^{2}|\!\Delta|^{2}\right\rangle _{x}\!=\!\frac{\pi^{2}}{%
4L^{2}\xi_{\mathrm{GL}}^{2}}\frac{\cos^{2}\frac{\pi\delta}{L}}{\cosh^{2} 
\frac{\pi y}{L}}\left( y^{2}\!+\!y_{0}^{2}\!+\!\frac{32}{27}\Gamma^{2}\xi_{%
\mathrm{GL}}^{2}\right) ,  \label{deltaphi}
\end{equation}

\begin{align}
& \left\langle \frac{\partial|\Delta|^{2}}{\partial x}\right\rangle _{x}
=\left( \frac{\pi}{2L}\right) ^{2}\frac{\Delta_{0}^{2}\left( \Gamma\right) }{%
\xi_{\mathrm{GL}}^{2}\cosh^{2}\frac{\pi y}{L}}\left( y^{2}+y_{0}^{2}+\frac{32%
}{27}\Gamma^{2}\xi_{\mathrm{GL}}^{2}\right)  \notag \\
& \cdot\left[ \frac{1}{2}\exp\left( -\frac{2\pi y}{L}\right) +\sin
^{2}\left( \frac{\pi\delta}{L}\right) \sinh\left( \frac{\pi y}{L}\right)
\exp\left( -\frac{\pi y}{L}\right) \right] ,  \label{delta1}
\end{align}

\begin{widetext}
\begin{align}
\left\langle \frac{\partial|\Delta|^{2}}{\partial y}\right\rangle _{x} &
=\frac{Z\left(  y,\Gamma\right)  \Delta_{0}^{2}\left(  \Gamma\right)  }%
{2L^{2}}\left\{  \frac{L^{2}}{4Z^{2}}\left(  \frac{\partial Z}{\partial
y}\right)  ^{2}\left[  1+\tanh^{2}\frac{\pi y}{L}\right]  +\frac{\pi L}%
{Z}\left(  \frac{\partial Z}{\partial y}\right)  \frac{\tanh\frac{\pi y}{L}%
}{\cosh^{2}\frac{\pi y}{L}}\right. \nonumber\\
&  \left.  +\pi^{2}\frac{\tanh\left(  \frac{\pi y}{L}\right)  }{\cosh
^{4}\left(  \frac{\pi y}{L}\right)  }\left(  \frac{3}{4}+\exp\left(
-\frac{2\pi y}{L}\right)  +\frac{1}{4}\exp\left(  -\frac{4\pi y}{L}\right)
\right)  \right\} \;y>0. \label{deltay}
\end{align}
\end{widetext}
In order to obtain the value of the activation energy from Eq. (\ref{GLfull}%
) one has to learn how to integrate the Eqs. (\ref{deltaphi})-(\ref{delta1})
over $y.$ We demonstrate here some of them: 
\begin{equation}
\int_{0}^{\infty}\frac{dy}{Z\left( y,\Gamma\right) }\left( \frac{\partial Z}{%
\partial y}\right) ^{2}=\frac{4}{3\xi_{\mathrm{GL}}}-\frac{\pi}{2\xi_{%
\mathrm{GL}}^{2}}\sqrt{y_{0}^{2}+\frac{32}{27}\Gamma^{2}\xi _{\mathrm{GL}%
}^{2}},
\end{equation}
\begin{widetext}
\[
\int_{0}^{\infty}\frac{dy}{Z\left(  y,\Gamma\right)  \cosh^{2}\frac{\pi y}{L}%
}\left(  \frac{\partial Z}{\partial y}\right)  ^{2}=\frac{L}{\pi
\xi_{\mathrm{GL}}^{2}}\left[  1-\frac{\pi^{2}}{L^{2}}\left(  y_{0}^{2}%
+\frac{32}{27}\Gamma^{2}\xi_{\mathrm{GL}}^{2}\right)  I_{2}\left(  \frac{\pi
}{2L}\sqrt{y_{0}^{2}+\frac{32}{27}\Gamma^{2}\xi_{\mathrm{GL}}^{2}}\right)
\right]  ,
\]
\begin{align*}
\int_{0}^{\infty}\left\langle \frac{\partial|\Delta|^{2}}{\partial
x}\right\rangle _{x}dy &  =\frac{\Delta_{0}^{2}\left(  \Gamma\right)  }%
{8}\frac{L}{\pi\xi_{\mathrm{GL}}^{2}}\left\{  \frac{3}{4}\zeta\left(
3\right)  -\frac{1}{2}\zeta\left(  2\right)  +\frac{\pi^{2}}{L^{2}}\left(
2\ln2-1\right)  \left(  y_{0}^{2}+\frac{32}{27}\Gamma^{2}\xi_{\mathrm{GL}}%
^{2}\right)  \right.  \\
&  \left.  +\sin^{2}\left(  \pi\delta/L\right)  \left[  \zeta\left(  2\right)
-\frac{3}{4}\zeta\left(  3\right)  +\frac{2\pi^{2}}{L^{2}}\left(
1-\ln2\right)  \left(  y_{0}^{2}+\frac{32}{27}\Gamma^{2}\xi_{\mathrm{GL}}%
^{2}\right)  \right]  \right\}  ,
\end{align*}%
\[
I_{1}\left(  \!\alpha\right)  =\int_{0}^{\infty}\frac{dx}{\cosh x}\frac
{1}{x^{2}+4\alpha^{2}}=\frac{\pi}{2}\left[  \frac{\pi}{2\alpha\cos2\alpha
}+2\sum_{n=0}^{\infty}\frac{\left(  -1\right)  ^{n}}{4\alpha^{2}-\pi
^{2}\left(  n+\frac{1}{2}\right)  ^{2}}\right]  ,
\]%
\[
I_{2}\left(  \!\alpha\right)  =\int_{0}^{\infty}\frac{dx}{\cosh^{2}x}\frac
{1}{\left(  x^{2}+4\alpha^{2}\right)  }=\frac{1}{2}\left\{  \frac{\pi}%
{2\alpha\cos^{2}2\alpha}-\sum_{n=0}^{\infty}\frac{4\pi^{2}\left(  n+\frac
{1}{2}\right)  }{4\alpha^{2}-\pi^{2}\left(  n+\frac{1}{2}\right)  ^{2}%
}\right\}  ,
\]
$\;\alpha>0.$
\end{widetext}

\section{}

Let us notice that if the function $\widetilde{\Delta }\left( x,y\right) $
is that one, for which the conditional extremum of the GL functional Eq. (%
\ref{GLfull}) is reached, the value of free energy in this state takes a
specially simple form:%
\begin{equation}
F_{s}\!=\!-\!\nu \int d^{3}\mathbf{r}\left[ \frac{7\zeta \left( 3\right) }{%
16\pi ^{2}T^{2}}|\Delta \left( \mathbf{r}\right) |^{4}\!+\!\frac{1}{c}\left( 
\mathbf{A}\!-\!\frac{c}{2e}\mathbf{\nabla }\varphi \right) \cdot \mathbf{j}%
_{\infty }\right] .  \label{FS}
\end{equation}%
This expression enables us to determine the value of parameter $r_{1}.$ In
order to do this we calculate the value of the GL\ free energy Eq. (\ref{FS}%
) using Eqs. (\ref{phl})-(\ref{dl}). In result one finds the equation%
\begin{equation}
\ln ^{2}\left( \frac{2\gamma L^{2}}{r_{1}^{2}}\right) \frac{\tanh Y}{\cosh
^{2}Y-\left( 3\mathcal{L}-2\right) /\mathcal{L}}=\mathrm{const,}
\label{const}
\end{equation}%
where $Y=y_{0}\sqrt{3\mathcal{L}-2}/\left( 2\xi _{\mathrm{GL}}\right) $ and
the value of the constant is independent on $y_{0}.$ The maximal value
of $r_{1}$ is reached when $Y$ satisfies the condition of extremum:%
\begin{equation}
\left\{ \frac{\partial }{\partial Y}\left[ \frac{\tanh Y}{\cosh ^{2}Y-\left(
3\mathcal{L}-2\right) /\mathcal{L}}\right] \right\} _{Y=Y_{\mathcal{L}}}=%
\mathrm{0.}  \label{ext}
\end{equation}%
Eq. (\ref{ext}) can be solved:%
\begin{equation}
\tanh Y_{\mathcal{L}}=2\left[ \frac{1-\mathcal{L}}{4-3\mathcal{L}+\sqrt{%
\mathcal{L}\left( 16-15\mathcal{L}\right) }}\right] ^{1/2}.
\end{equation}%
One can see, that our assumption that the in Eq. (\ref{const}) the value of the constant is independent on $y_{0}$ is confirmed 
(the found value $Y_{%
\mathcal{L}}$ is independent on it). Now we can use the found
value 
\begin{equation*}
y_{0}=\frac{2\xi _{\mathrm{GL}}}{\sqrt{3\mathcal{L}-2}}\mathrm{arctanh}
\left[ \frac{4\left( 1-\mathcal{L}\right) }{4-3\mathcal{L}+\sqrt{\mathcal{L}\left( 16-15%
\mathcal{L}\right) }}\right] ^{1/2}
\end{equation*}%
to perform the final integration in Eq. (\ref{FS}), what results in%
\begin{widetext}
\begin{align*}
\delta F_{s}  &  =4\nu\Delta_{0}^{2}\left(  \Gamma\right)  \tau S\xi
_{\mathrm{GL}}\sqrt{3\mathcal{L}-2}\left\{  \frac{1- \tanh Y_{\mathcal{L}}
}{6\mathcal{L}}\left[  4+\left(  3\mathcal{L}-2\right)  \tanh Y_{\mathcal{L}}\left(
1+\tanh Y_{\mathcal{L}}\right)  \right]  \right. \\
&  \left.  -\sqrt{\frac{2\left(  1-\mathcal{L}\right)  }{3\mathcal{L}-2}%
}\left[  \arctan\sqrt{\frac{3\mathcal{L}-2}{2\left(  1-\mathcal{L}%
\right)  }}-\arctan\left(  \sqrt{\frac{3\mathcal{L}-2}{2\left(
1-\mathcal{L}\right)  }}\tanh Y_{\mathcal{L}}\right)  \right]  \right\}  .
\end{align*}
\end{widetext}


\begin{references}
	\bibitem{S09} M.Sahu {\em et al,}   {\em Nature Physics} {\bf5}%
	, 503 (2009) and references therein.
	\bibitem{B13} A.Bezryadin   {\em Superconductivity in Nanowires}, Willey-VCH (2013).
	\bibitem{LA67}  J.S.Langer and Vinay Ambegaokar, {\em Phys. Rev.} {\bf
		164}, 498 (1967).
	\bibitem{T75}  Michael Tinkham, {\em Introduction to Superconductivity}%
	, McGraw-Hill Book Company, (1975).
	\bibitem{A88}  A.A.Abrikosov, {\em Fundamentals of the Theory of Metals}%
	, North Holland, (1988).
	\bibitem{LV04}  A.I.Larkin, A.A.Varlamov, {\em Theory of fluctuations in
		superconductors}, Oxford University Press, (2005).
	\bibitem{G59} L.P.Gor'kov, {\em Soviet Physics - JETP}, {\bf9}, 1364
	(1959).
	\bibitem{LO84} A.I.Larkin,  Yu.N.Ovchinnikov, {\em Soviet Physics - JETP},
	{\b59}, 420 (1984).
	\bibitem{MH70} D. E. MCCumber and B.I. Halperin {\em Phys. Rev.} {\bf
		B 1}, 1050 (1970).
	\bibitem{GE68} L.P.Gorkov and G.M.Eliashberg, {\em Soviet Physics - JETP}, {\bf27}, 328 (1968).
	\bibitem{J62}  B.D.Josephson, {\em Advances in Physics}, {\bf14}, 419 (1965).
	\bibitem{LO86} A.I.Larkin,  Yu.N.Ovchinnikov in
	{\em Nonequilibrium Superconductivity}, Edited by D.N.~Langenberg and A.I.Larkin.
	Elsevier Science Publ. B. V. (1986).
\bibitem{B12} T. Aref, A. Levchenko, V. Vakaryuk, and A. Bezryadin, {\em Phys. Rev.} {\bf B 86}, 024507 (2012).
\bibitem{B14} A. Murphy, A. Semenov, A. Korneev, Yu. Korneeva, G. Gol'tsman, A. Bezryadin.  arXiv:1410.7689 (2014)
\bibitem{V12} D. Y. Vodolazov, {\em Phys. Rev.} {\bf B 85}, 174507 (2012).
\end{references}
\end{document}